\documentclass[sigconf]{acmart}

\pdfoutput=1

\AtBeginDocument{%
  \providecommand\BibTeX{{%
    \normalfont B\kern-0.5em{\scshape i\kern-0.25em b}\kern-0.8em\TeX}}}

\acmConference[CHI '21]{ACM Conference on Human Factors in Computing Systems}{May 08--13, 2021}{Online Virtual}

\usepackage{xcolor}
\usepackage{paralist}
\usepackage{float}
\usepackage{placeins}
\usepackage{array}
\usepackage{textcase}

\definecolor{class}{RGB}{117, 161, 210}

\usepackage{multirow}

\usepackage{amsmath}

\usepackage{tikz}
\definecolor{nodeBG}{RGB}{29,40,97}
\newcommand*\circled[1]{\tikz[baseline=(char.base)]{
            \node[shape=circle,draw, fill=nodeBG,inner sep=0.5pt] (char) {\textcolor{white}{\textbf{#1}}};}}

\newcommand{\eg}{\textit{e.g.}}
\newcommand{\ie}{\textit{i.e.}}
\newcommand{\etal}{\textit{et al.}}
\newcommand{\name}{LQ\textsuperscript{2}}
\newcommand{\fullname}{Layout Quality Quantifier}
\newcommand{\MTurk}{MTurk}
\newcommand{\finalAcc}{78\%}

\newenvironment{rev}[0]{%
    \leavevmode\color{black}\ignorespaces
}{}

\acmSubmissionID{2322}

\begin{document}

\title{Learning to Automate Chart Layout Configurations \texorpdfstring{\\ Using Crowdsourced Paired Comparison}{}}
\renewcommand{\shorttitle}{Learning to Automate Chart Layout Configurations Using Crowdsourced Paired Comparison}


\author{Aoyu Wu}
\affiliation{Hong Kong University of Science and Technology}
\email{awuac@connect.ust.hk}

\author{Liwenhan Xie}
\affiliation{Hong Kong University of Science and Technology}
\email{liwenhan.xie@connect.ust.hk}

\author{Bongshin Lee}
\affiliation{Microsoft Research}
\email{bongshin@microsoft.com}

\author{Yun Wang}
\affiliation{Microsoft Research Asia}
\email{wangyun@microsoft.com}

\author{Weiwei Cui}
\affiliation{Microsoft Research Asia}
\email{Weiwei.Cui@microsoft.com}

\author{Huamin Qu}
\affiliation{Hong Kong University of Science and Technology}
\email{huamin@cse.ust.hk}
\renewcommand{\shortauthors}{Wu, et al.}

\begin{abstract}
We contribute a method to automate parameter configurations for chart layouts by learning from human preferences. 
Existing charting tools usually determine the layout parameters using predefined heuristics, producing sub-optimal layouts. 
People can repeatedly adjust multiple parameters (\eg{}, chart size, gap) to achieve visually appealing layouts. 
However, this trial-and-error process is unsystematic and time-consuming, without a guarantee of improvement. 
To address this issue, we develop Layout Quality Quantifier (\name{}), a machine learning model that learns to score chart layouts from paired crowdsourcing data. 
Combined with optimization techniques, \name{} recommends layout parameters that improve the charts’ layout quality. 
We apply \name{} on bar charts and conduct user studies to evaluate its effectiveness by examining the quality of layouts it produces.
Results show that \name{} can generate more visually appealing layouts than both laypeople and baselines. 
This work demonstrates the feasibility and usages of quantifying human preferences and aesthetics for chart layouts.
\end{abstract}

\ccsdesc[500]{Human-centered computing~Visualization design and evaluation methods}
\ccsdesc[100]{Human-centered computing~Visualization toolkits}


\keywords{Machine Learning, Visualization, Crowdsourced, Visual Design, Image Quality Assessment}


\maketitle

\section{Introduction}
Data visualizations have been ubiquitous in everyday life, such as social media, magazines, and websites.
They are widely used by the general public to express complex data in an intuitive, concise, and visually appealing manner.
However, creating effective and elegant visualizations is a challenging task even for professions~\cite{qin2019making}.
Individuals usually need to engage in a time-consuming process to craft designs that clearly convey information and insights, 
while satisfying the aesthetic goals. 
As such, there have been huge efforts from both industry and research communities to aid the design process by automated approaches.

\begin{figure}[!t]
	\centering
	\includegraphics[width=1\linewidth]{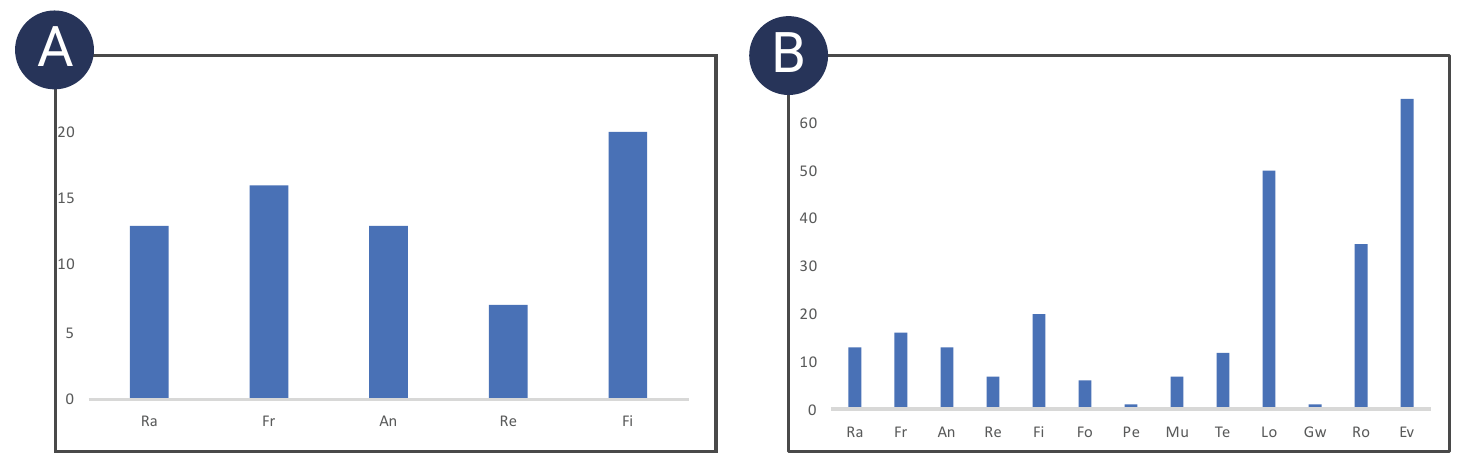}
	\caption{Visualization tools such as Microsoft Excel utilize a default heuristic to generate layouts: {\protect\circled{A}}~the chart with five bars; {\protect\circled{B}}~increasing the number of bars results into a chart with the same size but different bandwidths. Those layouts have room for improvements through manual refinement.}
    \Description[Charting tools might generate sub-optimal layouts.]{Existing charting tools such as Microsoft Excel utilize default heuristics to generate layouts, which might become sub-optimal.}
	\label{fig:movEx}
\end{figure}

Existing approaches have predominantly focused on studying and optimizing performance metrics for data analytics concerning usability and utility.
For example, commercial software such as Excel automatically recommends chart types based on selected data.
Besides, much recent research proposes automated visualization systems that retain data integrity~\cite{hopkins2020visualint},
highlight interesting data facts~\cite{lai2020automatic},
and recommend effective visual encodings~\cite{moritz2018formalizing,dibia2019data2vis, hu2019vizml}.
Nevertheless, those systems utilize pre-defined heuristics to generate visual styles, which could be sub-optimal (\autoref{fig:movEx}).
This paradigm results in a quasi-automated process where individuals need to manually adjust the visual style of the automatically generated charts (\eg{},~\cite{Tang2020PlotThread, wang2019datashot}).
However, performing manual adjustments can be unsystematic and difficult, especially for lay users without design backgrounds~\cite{Ye20Easy}.
Users might be unaware of guidance or find it tedious to adjust multiple parameters simultaneously.
To address this problem, we aim to propose a systematic data-driven approach that recommends parameter configurations by learning from crowdsourcing human preference data.
Particularly, we study layouts because it is a fundamental element of chart design~\cite{Tableau:Best}.
We focus on bar charts which are one of the most common chart types~\cite{battle2018beagle}.

Despite the increasing acknowledgement of the importance of visual styles in charts~\cite{cawthon2007effect,harrison2015infographic,kennedy2018feeling,reppa2015going,moere2012evaluating},
little work has attempted to understand and quantify layout qualities through large-scale user studies.
Research in graphical design has provided various layout metrics such as alignment and segmentation~\cite{o2014learning},
but they are not readily applicable to charts that are data-driven and yield different visual perception~\cite{bylinskii2017learning, Vaca:2018}.
There is a lack of empirical studies to understand metrics for chart layouts.
This is challenging due to the subjective nature of layout qualities,
which requires a large number of participants to score charts.
The scores, however, might not be precise since participants might be hesitant and feel difficult to give an accurate score~\cite{rouse2010tradeoffs, tsukida2011analyze}.
Besides, the scoring scales can be inconsistent among participants~\cite{gao2015learning}.
Those limitations constrain the reliability of utilizing the scores as benchmarks for machine learning models.
To that end, our approach is inspired by the successful applications of pair-wise ranking for assessing natural image qualities~\cite{kong2016photo}. 
We propose a two-alternative forced-choice experiment~\cite{fechner1860elemente}, asking participants to select a better chart layout between two candidates. 
This data acquisition method allows us to obtain more precise and consistent results~\cite{rouse2010tradeoffs, bradley1952rank}.

We propose a novel approach, called \fullname{} (\name{}), for learning to score and rank the chart layout configurations from human preference through crowdsourced pair-comparison experiments.
\name{} utilizes neural networks to predict the score of an individual chart by taking comparison pairs as training data.
\name{} predicts the pair-wise ranking with the accuracy of \finalAcc{}, 
showing that it could reasonably learn human preference for layout configurations.
We further interpret the trained model by investigating the impact of layout parameters on human preference,
thereby summarizing rule-of-thumb for layout configurations in bar charts.
Finally, quantitative user studies demonstrate that \name{} could recommend more visually appealing layouts than manual results by laypeople and default styles in Excel and Vega-Lite~\cite{satyanarayan2016vega}.
Overall, our work demonstrates the possibility of quantifying human aesthetics for charts.
We open source all our code and experimental material\footnote{\url{https://github.com/shellywhen/LQ2}}.
In summary, our contributions are as follows:
\begin{compactitem}
    \item A novel approach for quantifying human preference for chart layouts through crowdsourced paired comparison
    \item A machine-learning method, \name{}, for ranking and scoring layout configurations in bar charts
    \item A set of qualitative and quantitative evaluations as well as two user studies that demonstrate the effectiveness and usefulness of \name{}
\end{compactitem}



\section{Related Work}
Our work is related to aesthetics for visualizations, automated visualization designs, as well as data collection and training for visualization research. 

\subsection{Aesthetics for Visualizations}
In a broader sense, our work is related to the aesthetic qualities of data visualizations.
In the book \textit{Information is Beautiful}, McCandless~\cite{McCandless09} lists aesthetics as one of the four criteria for a good visualization.
However, aesthetics were traditionally considered as an add-on feature that was typically implemented at the very end of the design process.
Already 13 years ago, Cawthon and Moere~\cite{cawthon2007effect} argued for increased recognition for visualization aesthetics,
by demonstrating the relationships between aesthetics and usability in data visualizations. 
Since then, many empirical studies have shown that the aesthetics of data visualizations could contribute to various factors such as first impressions~\cite{harrison2015infographic}, memorability~\cite{borkin2013makes}, emotional engagement~\cite{kennedy2018feeling}, and task performances~\cite{reppa2015going}.

Nevertheless, little work has studied what makes a data visualization visually appealing.
Moere \etal{}~\cite{moere2012evaluating} demonstrated that the visual styles could lead to different comments regarding aesthetics.
Quispel \etal{}~\cite{quispel2016graph} found that laypeople were attracted to designs they perceived as familiar and easy to use.
However, they investigate aesthetics as a qualitative reflection of personal judgment rather than a quantifiable and comparable entity.
Human preferences for aesthetics in the context of charts are still not methodically quantifiable from a data-driven perspective, and seem underrepresented in large-scale empirical studies.
We address this gap by proposing a systematic machine learning approach for ranking and scoring layout qualities from crowdsourcing experiment data.
Besides, we propose our approach for interpreting the trained model, 
whereby summarizing speculative hypotheses that warrant future empirical research to confirm.

\subsection{Automated Visualization Design}
Recently, there have been growing interests in applying machine learning methods for automated visualization designs.
Researchers have proposed many systems~\cite{hu2019vizml, dibia2019data2vis, moritz2018formalizing, luo2018deepeye} that recommend visualizations based on data structures and characteristics.
Those systems focus on deciding the effective chart type, visual encoding, and data transformation.
In addition to effectiveness, much research has been devoted to optimizing visualizations from other aspects. 
For example, VisuaLint~\cite{hopkins2020visualint} addresses the data integrity by surfacing chart construction errors such as truncated axes.
DataShot~\cite{wang2019datashot} and Calliope~\cite{shi2020calliope} focus on generating visualizations with interesting data-related facts from tabular data.
Dziban~\cite{lin2020dziban} attempts to balance automated suggestions with user intent by preserving similarities with anchored visualizations.
Different from them, our work studies to improve the aesthetic quality of chart layouts.

\begin{figure*}[!t]
	\centering
	\includegraphics[width=1\linewidth]{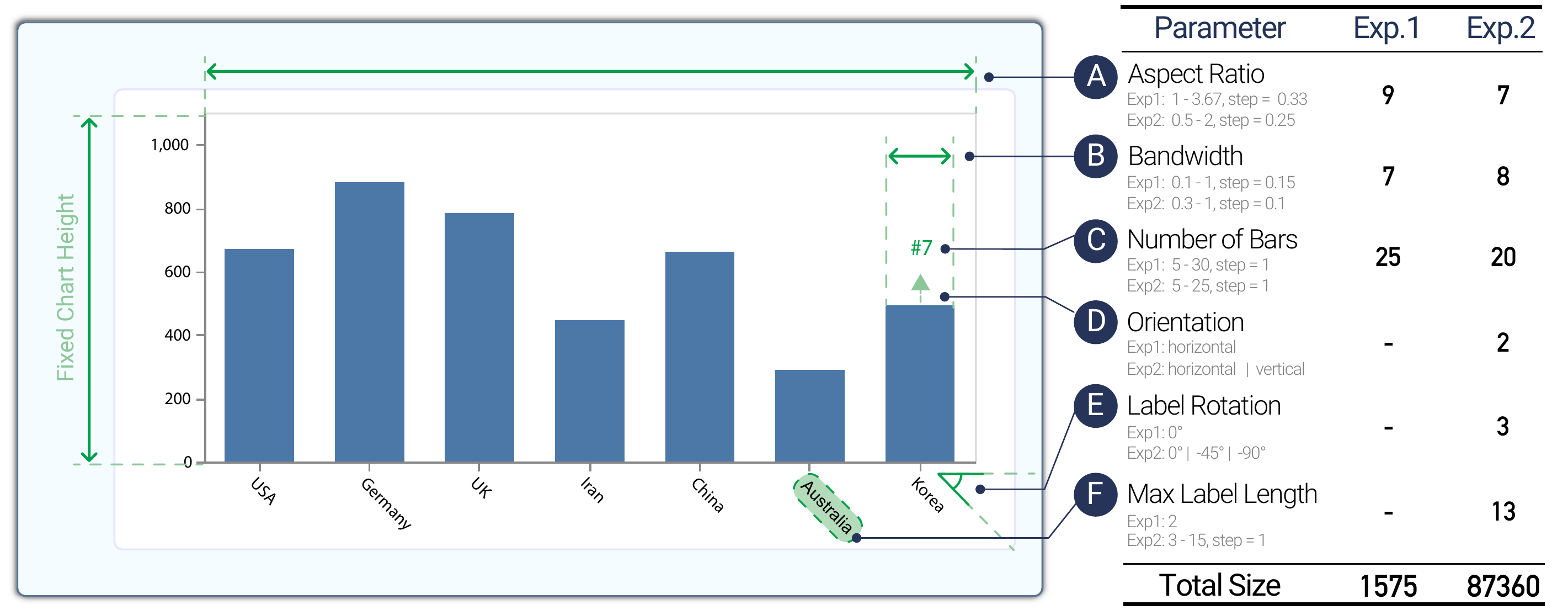}
	\caption{The layout parameters in Experiment 1 and 2. The table displays the sampled values of each parameter. Experiment 1 concerns 3 parameters with 1,575 possible combinations, while Experiment 2 includes 6 parameters with 87,360 combinations.}
	\label{fig:parameter}
    \Description[Our experiments study six layout parameters]{Our experiments study six layout parameters with 1,575 and 87,360 possible combinations of parameter values.}
\end{figure*}

Researchers have also recently proposed many approaches for improving the visualization layout. 
Several work automatically extracts reusable layout templates from visualizations~\cite{chen2019towards,chen2020augmenting} and infographics~\cite{lu2020exploring}.
Nevertheless, they do not propose metrics for extracted layouts.
Other systems optimize the layout according to various metrics such as mobile-friendliness~\cite{Wu2020MVF}, similarities with user-input layouts~\cite{Tang2020PlotThread} and graph features~\cite{haleem2019evaluating,wang2019deepdrawing}.
However, their metrics are not derived from empirical studies and therefore might not reflect the overall perceived quality~\cite{behrisch2018quality}.
Therefore, our work studies how to quantify and optimize layout qualities from crowdsourcing experiments.


\subsection{Data Collection and Training for Visualization Research}
Recent years have witnessed a growing recognition of data collection to facilitate research in machine learning for visualizations. 
Many efforts have been made to collect real datasets of charts from websites~\cite{battle2018beagle, hu2019vizml, poco2017reverse} and scientific literature~\cite{lee2017viziometrics, chen2020composition}.
Those datasets include annotations or original data as ground-truth labels and assume that those charts share the same quality.
Therefore, another line of research conducts crowdsourcing user studies to obtain quality metrics such as task completion time and accuracy~\cite{saket2018task} as well as attentions~\cite{kim2017bubbleview}, which can be measured objectively by devices.
However, it is much more challenging to generate a reliable dataset for subjective metrics, as participants might not share a reliable and consistent scoring scale~\cite{rouse2010tradeoffs, tsukida2011analyze, gao2015learning}.
\begin{rev}
To that end, 
Saket \etal{}~\cite{saket2018task} extended their experiment by asking participants to rank five different visualization types in the order of preference and found a positive correlation between user preference and task accuracy.
However, the collected data is for statistical analysis instead of machine learning tasks.
\end{rev}

To generate dataset for machine learning, 
Luo \etal{}~\cite{luo2018deepeye} proposes a pair-wise comparison approach, \ie{}, to ask participants to choose which chart is better from two candidates, which yields more precise results.
They subsequently compute the overall order from pair-wise comparison, and choose the top-ranked ones as training data.
However, their approach is limited for two reasons.
First, it is inefficient as they only obtained 2,520/30,892 good/bad charts after 285,236 comparisons.
Second, they formulate the problem as a classification task, neglecting the subtle differences among charts.
To that end, we propose~\name{} which predicts a numerical score of a single chart through regression neural networks, while directly taking paired comparisons as the training input.
\name{} is built on similar learning frameworks for image assessment~\cite{wang2014learning},
but integrates a parameter module and two sampling strategies to learn the visualization-specified features.

\section{Overview}
Data visualizations represent data with graphical elements according to visual specifications.
Specifications can be classified into two types: \textit{visual encodings} that map data to visual properties (\eg{}, color, position, size) of graphical elements, and \textit{visual styles} that specify the remaining visual properties irrespective of data (\eg{}, label rotation, bar bandwidth).
Our work aims to automate the parameter configurations for the latter, \ie{}, visual styles, 
which are largely neglected in existing automated charting tools.
Concretely, we focus on the layout properties in bar charts, 
since this is one of the first work that attempts to rank and recommend parameter specifications of visual styles leveraging machine-learning approaches.
In this section, we describe our experiments, the design considerations, and the problem formulation.

\subsection{Experiments}
\label{overview:experiments}
Different from visual encodings that are typically described as discrete mappings, visual styles usually have greater cardinality and continuous values that increase their complexity.
To keep this study complexity manageable, we select two concrete yet underexplored experiments (\autoref{fig:parameter}).

\begin{figure*}[t]
    \setlength{\belowcaptionskip}{-5pt}
	\centering
	\includegraphics[width=1\linewidth]{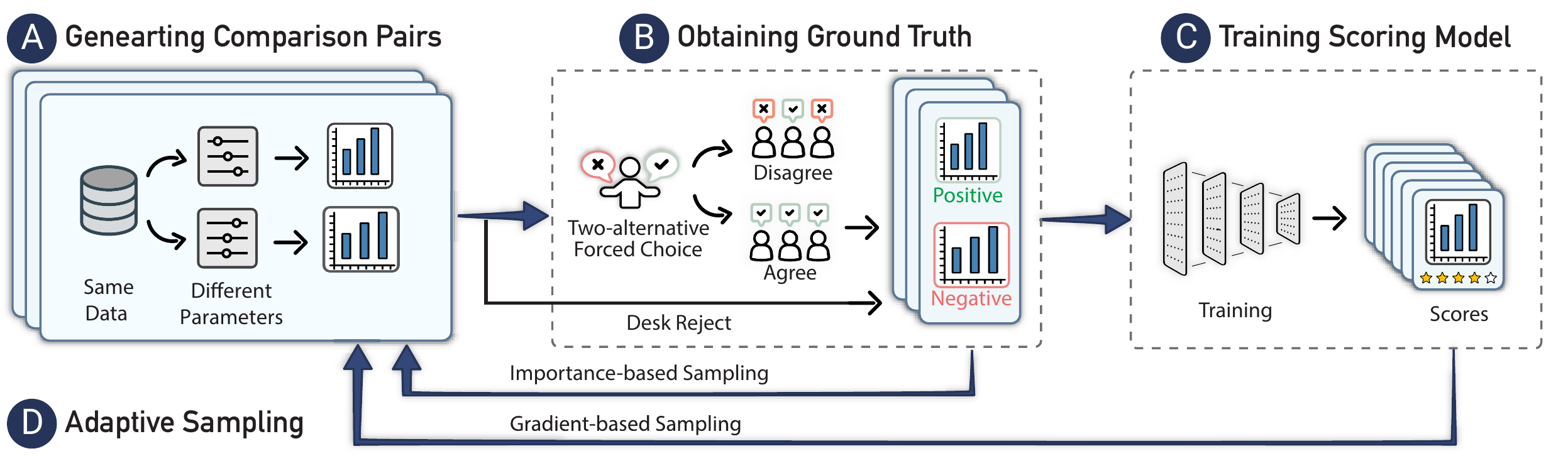}
	\caption{The data collection process in an iterative manner: {\protect\circled{A}} Generating paired charts with same data and different layout parameters; {\protect\circled{B}} Labelling the training data through crowdsourcing experiments; {\protect\circled{C}} Training the scoring models; {\protect\circled{D}} Utilizing two offline adaptive sampling strategies to increase the representativeness of the training dataset.}
	\label{fig:pipeline}
    \Description[Our design collection process contains four step]{Our design collection process contains four steps: generating comparison pairs, obtaining ground truth, training score models, and adaptive sampling.}
\end{figure*}

Our first experiment considers three basic layout-related parameters, \ie{}, the number of bars, the aspect ratio of the chart, and the bandwidth.
This is because we observe that existing charting tools determine those values by predefined heuristics.
For instance, Microsoft Excel fixes the aspect ratio and computes the bandwidth according to the number of bars (\autoref{fig:movEx}).
In this paper, we argue and demonstrate that such default heuristics could result in sub-optimal layouts. 
Individuals, therefore, need to repeatedly adjust several parameters to achieve visually appealing layouts.
However, such manual adjustments are unsystematic and time-consuming, without a guarantee of improvement.
Therefore, we study how to automatically configure those parameters by learning from human preferences.

The second experiment is extended with another three parameters, namely, the chart's orientation, the max length and rotation degree of axis tick labels.
This experiment is motivated by the practical needs for responsive visualization design,
\ie{}, how to adjust the chart layout to fit into different sizes.
This is a challenging task since chart creators need to manually examine and edit layouts for multiple chart sizes~\cite{hoffswell2020techniques}.
Therefore, we add those three parameters that are often subject to adjustments in responsive visualization designs.
This experiment extends existing automated responsive visualization approaches~\cite{Wu2020MVF} by considering the aspect ratio and allowing rotating chart orientations.

\subsection{Design Consideration}
The problems above can be summarized as optimization problems, \ie{}, to find values of visual styles that maximize the layout quality.
To guide the design of our solution, we summarize two primary considerations:

\paragraph{C1: To quantify and score layout qualities.}
One of the primary challenges in optimization problems is to define the objective function.
Hence, our primary goal is to learn a loss function that maps values of layout parameters onto a numerical score intuitively representing the layout qualities.
This loss function can be subsequently used for mathematical optimization.

\paragraph{C2: To learn the overall quality from human feedback.}
Since judgments of layout qualities involve a wide range of factors,
previous work in graphic designs~\cite{o2014learning, li2020attribute} usually utilizes human-crafted metrics (\eg{}, symmetry) to measure layout qualities.
Their methods face challenges in the context of data visualizations since few human-crafted metrics are available for chart layouts.
Besides, it is difficult to weigh different metrics to reflect the overall quality perceived by users.
Therefore, we aim to measure the overall quality by learning from human feedback,
and conduct post hoc analysis to summarize rule-of-thumb from the trained model.

\subsection{Problem Abstraction and Formulation}
Guided by the design considerations, 
our main task is to develop a machine-learning model that learns to predict a layout quality score of the given parameters from human feedback data.
We formulate this task as a learning-to-rank problem~\cite{liu2011learning}, which purposes to acquire a global ranking from partial orders.
The ground truth of partial orders is harvested from experimental data on human preference.
Specifically, we conduct a paired comparison experiment, asking participants to choose their preferred layout from two candidates.
The results from paired comparisons contain partial orders, which constitute the training data.

\section{Data Collection}
\label{data}

We describe our process of constructing the training dataset containing ranked pairs of chart layout configurations.
As shown in \autoref{fig:pipeline}, the process is iterative and contains four steps.
In this section, we describe the step \circled{A}, \circled{B} , and \circled{D} in \autoref{fig:pipeline} in detail.
Step \circled{C} will be introduced in \autoref{model}.


\subsection{Generating Comparison Pairs}
\label{dataset:generatePairs}
Our first step is to create paired charts for crowdsourcing comparison.
We decide to synthesize charts since it is difficult to harvest real-world chart pairs that are fairly comparable, that is, they are controlled to represent the same data.
Charts are created using Vega-Lite~\cite{satyanarayan2015reactive}, which allows specifying the aforementioned parameters in a declarative manner.
For each pair, we choose data from two popular real-world datasets, namely the Car dataset\footnote{\url{https://vega.github.io/vega-datasets/data/cars.json}} and the Baseball dataset\footnote{\url{https://github.com/vincentarelbundock/Rdatasets/blob/e38552ac3cb40a532941b09d7332b03d19409919/doc/plyr/baseball.html}}, 
and randomly select entries according to the number of bars.
In Experiment 1, we replace the tick labels with meaningless two-character tokens.
In Experiment 2, the tick labels are truncated according to the parameter of label lengths.

The remaining parameters are generated with different values within a chart pair, including the aspect ratio, chart orientation, bandwidth, and rotation of axis labels.
It is expensive to conduct controlled experiments for each parameter, since those parameters may not be interdependent.
Therefore, we choose to randomize all parameters, intuitively intending to obtain a wide variety of chart configurations.
However, exhaustive enumeration of possible values and combinations of parameters is infeasible due to their continuous distributions.
Thus, we decide to randomly sample from uniform distributed values (\eg{}, the bandwidths range from 0.1 to 1.0 with a step of 0.15).
We choose a relatively large step in order to make the differences notable.
While parameters are sampled randomly,
we make the sampled values evenly distributed to avoid the data imbalance problem.

\autoref{fig:parameter} shows the sample values and the possible combinations of parameter values in our experiments.
We update the sampling values in Experiment 2 according to our findings in Experiment 1.
For instance, we truncate the maximal aspect ratio to 2, since we observe that larger aspect ratios are less favored. 
It should be noted that the resulting design space is still considerably large that poses challenges in solving the optimization problem.

\subsection{Obtaining Ground Truth}
\label{data:crowdsourcing}
We harvest ground truths of ranked pairs of charts through a two-step process.

\begin{figure}[!b]
	\centering
	\includegraphics[width=1\columnwidth]{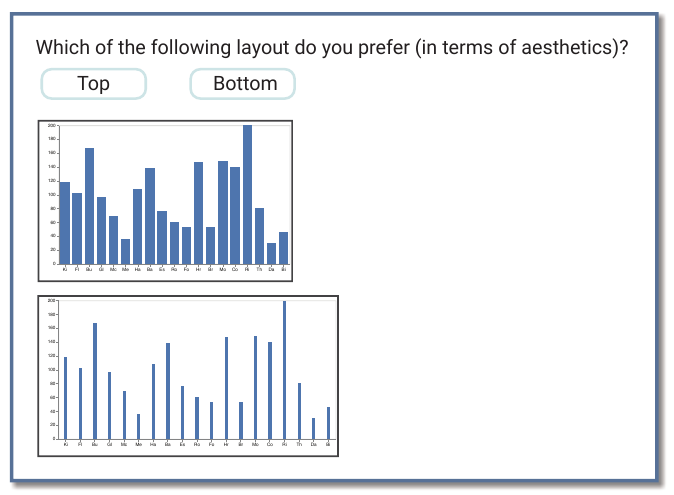}
	\caption{Illustration of the \MTurk{} interface for crowdsourcing experiments: we propose a two-alternative forced-choice design that makes it easier for participants to evaluate the relative quality of paired charts than scoring a single chart.}
	\label{fig:MTurkInterface}
    \Description[The crowdsourcing interface]{Our crowdsourcing interface asks participants to select the preferred layout from two choices.}
\end{figure}

\paragraph{Desk Reject}
First, we ``desk reject'' charts that violate a set of predefined rules and label them as negative in a pair.
We discard a chart pair if both charts violate rules.
Specifically, two rules are included in Experiment 2: the axis labels should not overlap with each other, and the axis label should not rotate in a horizontal bar chart.
This approach allows us to train an ML model that learns human-crafted rules during the training time and therefore better reflect the overall quality.
An alternative approach in visualization recommendation systems is to utilize rules as hard constraints in the optimization phase~\cite{moritz2018formalizing},
which, however, poses challenges in solving the optimization problem with the increasing number of rules.
This might be undesirable since it could prolong the execution time and therefore degrade the usability. 

\paragraph{Crowdsouring Experiments}
Second, we conduct crowdsourcing experiments on Amazon Mechanical Turk (\MTurk{}) to obtain experimental data from human preference.
\autoref{fig:MTurkInterface} illustrates the settings of the \MTurk{} experiment.
We propose a two-alternative forced-choice (2AFC) experiment, asking participants to choose ``which of the following two layouts do you prefer, in terms of aesthetics?''.
Two charts are placed vertically within a viewport since it is easy to compare by moving eyes between side-by-side views~\cite{munzner2014visualization}.
We choose a forced-choice method in an attempt to capture the subtle differences~\cite{zhao2018characterizes}.
Each \MTurk{} HIT consists of 10 comparison tasks, and each task (paired comparison) is assigned to 3 participants. 
For quality control, we randomly duplicate one comparison task within a HIT and swap the order of paired charts.
We keep HITs where participants offer consistent answers for duplicated tasks.

We measure the inter-observer reliability by the joint probability of agreement.
It is observed that three participants make the same choices in 45.6\% of pairs for two experiments.
This observation probability is much higher than that of the agreement by chance, \ie{}, 25\%,
showing that human preference exhibits a fair degree of agreement on layout qualities.
This fair agreement can have several reasons.
First, the differences between the two charts in a pair might be small and therefore cause uncertainties,
since the layout parameters are generated randomly.
Second, individual participants have different preferences.
Third, it might be because of the noises of \MTurk{} experiments.

We select paired comparisons with full agreements among participants as the training data to reduce noises~\cite{zhao2018characterizes}.
Each pair consists of two charts, denoted $\langle \mathcal{I^+}, \mathcal{I^-} \rangle$,
where $\mathcal{I^+}$ is preferred over $\mathcal{I^-}$.

\subsection{Adaptive Sampling}
It is crucial to employ an effective pair sampling strategy to select the most important pairs for rank learning~\cite{wang2014learning}.
Our uniform sampling and random pairing strategy in \autoref{dataset:generatePairs} is sub-optimal, 
since we are interested in finding the most ``optimal'' chart configurations.
Therefore, we propose two offline adaptive sampling strategies to improve the qualities and representativeness of the comparison pairs.
The term ``offline'' here is referred in the context of machine learning, 
that is, we re-sample comparison pairs when the initial training phase has finished.

\paragraph{Importance-based Sampling}
We are interested in finding important pairs that allow us to determine the ``best'' chart configurations.
Therefore, we borrow the idea of elimination tournaments, intuitively conducting a second round of comparisons among previous winners.
However, this is not readily applicable since our sample size is much smaller than the huge number of possible parameter combinations.
Thus, we propose an important-based sampling scheme, which intends to increases the probability of sampling important charts with ``good'' parameter values.

Suppose each chart $\mathcal{I}$ is configured by a set of parameters $p_i \in \mathcal{P}$, where the possible values of $p_i$ are $v_i^j \in \mathcal{V}_i$.
Let $w_i^j$ denote that number of times that the chart $\mathcal{I}$ whose configuration contains $v_i^j$ has won in paired comparisons in \autoref{fig:pipeline}~{\protect\circled{B}}.
We update the probability of sampling the value:
\begin{equation}
\label{equ:importanceSampling}
P(v_i^j) = \dfrac{\min{\{w_i^j, T\}}}{\sum_{j} \min{\{w_i^j, T\}}}, 
\end{equation}
where $T$ is a parameter responding the exploration-exploitation trade-off by avoiding empty probabilities.

\begin{table*}[!t]
\caption{Comparison of the performances between our method and baseline approaches in terms of the prediction accuracy (\%) via Monte-Carlo Cross-Validation for 10 runs with an 80-20 training-testing split ratio.}
\label{table:results}
\begin{tabular}{lccccccc}
\toprule
                   & \textbf{Ours} & \textbf{RankSVM} & \textbf{White Space} & \textbf{Scale} & \textbf{Unity} & \textbf{Balance} & \textbf{All} \\ \hline
\textbf{Exp. 1} (N = 1,177) &  \underline{76.60}    &   70.83      & 57.28  &  56.26  & 52.00   &  56.08  &  60.81   \\
\textbf{Exp. 2} (N = 1,333) &  \underline{78.27}    &   64.48      & 58.24   & 61.72   &  56.21  & 63.18   & 68.73    \\ \bottomrule
\end{tabular}
\end{table*}

\paragraph{Gradient-based Sampling}
Having a large step size in uniformly sampling might cause the model to overlook a maximal.
To address this problem, we use a gradient-based sampling method to sample important parameter values with a smaller step size.
As gradients are computed on a differentiable function,
we refer to our scoring model trained in \autoref{fig:pipeline}~{\protect\circled{C}}.
This scoring model learns a regression function $f(\cdot)$ that maps the parameter vector $p=\{p_1, p_2...,p_i\}$ to a numerical score.
We compute the locations where the gradient of $f$ along with $p$, $\nabla f(p)$, is smaller than a given threshold.
We sample parameter values within those locations with a smaller step-size, \ie{}, 1/3 of the original step-size.

In both experiments, we conduct each of the following adaptive sampling strategies once, and merge the resulting dataset.
This procedure results in 1,177 pairs in Experiment 1 and 1,333 pairs in Experiment 2.
Overall, our data collection process involves 416 unique \MTurk{} participants.
\section{Method}
\label{model}
\begin{figure}[!b]
	\centering
	\includegraphics[width=1\columnwidth]{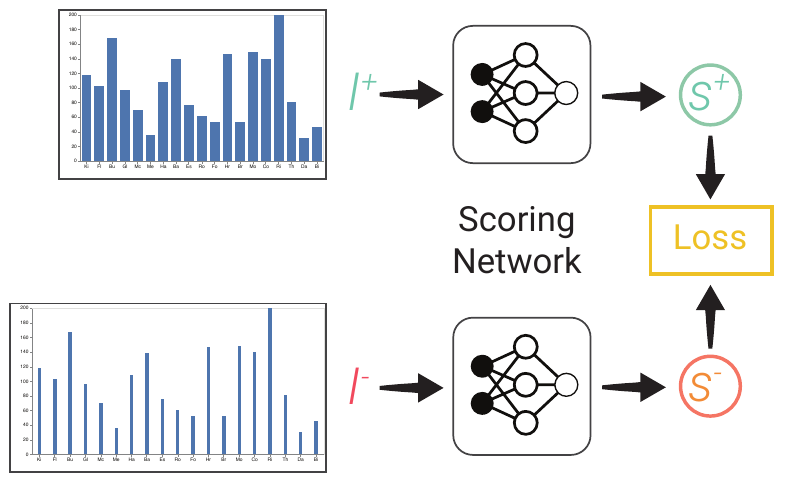}
	\caption{\name{} utilizes a Siamese neural network structure to work in tandem on a pair to compute comparable output.}
	\label{fig:model}
    \Description[The network structure]{Our network structure contains a Siamese neural network.}
\end{figure}

With the obtained pairs $\mathcal{D}=\{ \langle \mathcal{I^+}, \mathcal{I^-} \rangle \}$, \name{} aims to quantify the aesthetic scores of a given chart.
Specifically, we formulate the problem as a regression problem, 
that is, to output a numerical score $\mathcal{S}$ for an input chart $\mathcal{I}$.
Our goal is to learn a regression function $f(\cdot)$ that predicts a higher score for the preferred chart in a pair:
\begin{equation}
\label{equ:goal}
\mathcal{S^+} > \mathcal{S^-}, \forall \langle \mathcal{I^+}, \mathcal{I^-} \rangle \in \mathcal{D}
\end{equation}
where $\mathcal{S} = f(\mathcal{I})$.

\paragraph{Model Architecture}
\label{model:architecture}
\name{} adopts a Siamese neural network structure, 
\ie{}, to work in tandem on two different inputs with the same weights to compute comparable output~\cite{chopra2005learning}.
As shown in \autoref{fig:model},
it consists of two identical scoring networks,
and the loss function is defined on the combined output of scoring networks.
\begin{rev}
The scoring network takes the parameter values as input and outputs a numerical score.
We employ fully-connected neural networks (NN), which have proven effective in handling features describing design choices (\ie{}, parameters) in VizML~\cite{hu2019vizml}.
\end{rev}
Our NN contains 6 hidden layers,
each consisting of different numbers of neurons with ReLU activation functions and dropout layers.
We perform min-max normalization on the parameter values so that each parameter contributes approximately proportionately to the results.

\begin{rev}
We also tried to take the graphical features (\ie{}, images) as the training input with off-the-shelf Convolutional Neural Networks (CNNs) models.
However, this method did not bring about remarkable performances despite the expensive training time.
Our findings conform to earlier work~\cite{haehn2018evaluating,fu2019visualization} that CNNs might not readily capture human perception in visualizations.
Thus, we utilize the parameter as the input,
which serves as a compact and learning representation that reduces the computational costs.
\end{rev}





\paragraph{Loss Function}
We adopt the Pairwise Ranking Loss as the loss function, which explicitly exploits relative rankings of chart pairs~\cite{kong2016photo}:  
\begin{equation}
\label{equ:pairLoss}
\mathcal{L}(\mathcal{S^+}, \mathcal{S^-}) = max(0, \mathcal{S^+} - \mathcal{S^-} + m),
\end{equation}
where $m$ is a specified margin hyper-parameter.
This loss imposes a ranking constraint by penalizing mistakes for assigning a lower score to the preferred chart.

\paragraph{Implementation and Training}
We implement \name{} with Pytorch.
During training, we split the data by a ratio of 8:2 with the purpose of training and validation.
\begin{rev}
We tune several hyper-parameters by diagnosing the learning curves so that the plots of training and validation data converge to a good point of stability and have a small gap.
\end{rev}
The model is trained with the Adadelta optimizer for 200 epochs.
The learning rate is 1, and subsequently is reduced by half per 30 epochs.
We found that only the margin hyper-parameter $m$ had a significant impact on the training performance, 
while weight decay, optimizer, and dropouts had small effects.


\section{Experiment}
\label{experiment}
To evaluate the effectiveness of our method, we conduct experiments with baseline approaches and perform qualitative analyses with the trained scoring network across different layout parameters.

\begin{table*}[!t]
\caption{The Pearson correlation between predicted scores and each layout parameter in Experiment 1 and 2.}
\label{table:corr}
\begin{tabular}{lcccccc}
\toprule
             & \textbf{Number of Bars} & \textbf{Aspect Ratio} & \textbf{Bandwidth} & \textbf{Max Label Length} & \textbf{Label Rotation} & \textbf{Orientation} \\ \hline
\textbf{Exp. 1} & -0.38          & 0.20         & 0.27      &      -             &           -     &      -       \\
\textbf{Exp. 2} & -0.09          & 0.37         & -0.05     & -0.09             & -0.43          & 0.04   \\
                       \bottomrule         
\end{tabular}
\end{table*}

\begin{figure*}[!t]
	\centering
	\includegraphics[width=1\linewidth]{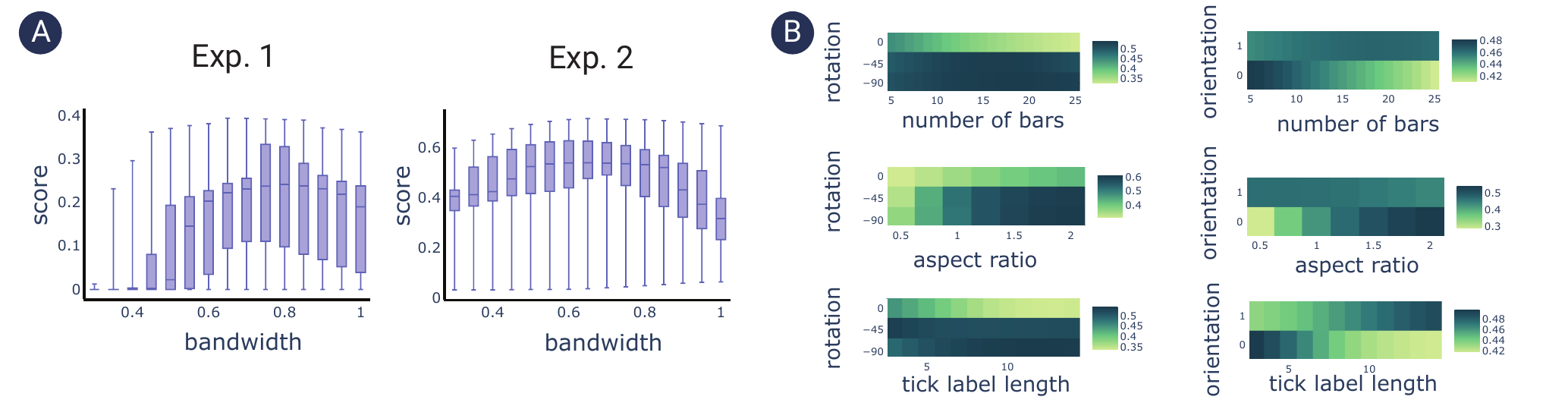}
	\caption{Visualizing the predicted scores with~{\protect\circled{A}} a single parameter by box-plots and ~{\protect\circled{B}} multiple parameters by heat-maps.}
	\label{fig:modelInterpret}
    \Description[Visualizing the predicted score]{Visualizing the predicted score. The box-plots shows that in Experiment 1, the the average scores are relatively higher when the bandwidth is between 0.6 and 0.95, with a subtle peak at 0.8. However, in Experiment 2, the ``optimal'' interval become between 0.5 to 0.85. The heat-maps show strong correlations between rotation and orientation with other parameters.}
\end{figure*}

\subsection{Performance}
We compare the performance of our model with several baseline approaches.
For experiment reproducibility, we adapt a Monte Carlo Cross-Validation strategy~\cite{xu2001monte},
that is, to randomly split the data into training data and testing data with an 80-20 ratio, run the experiment, and repeat the above process ten times.

\paragraph{Model Baseline}
Our problem is formed as a learning-to-rank problem.
Therefore, we consider the Ranking Support Vector Machine (\textbf{RankSVM})~\cite{joachims2002optimizing} as the baseline approach, which is a well-established method for computing the overall ranking based on pairwise preference.
Similar to Draco~\cite{moritz2018formalizing}, we use a linear SVM model with hinge loss.

\paragraph{Scoring Baseline}
We also compare our learned scoring network with existing human-crafted metrics for layout qualities in graphical design~\cite{o2014learning}.
We select four metrics that are mostly applicable to the context of charts, including \textbf{White Space}, \textbf{Scale}, \textbf{Unity}, and \textbf{Balance}.
Those metrics are implemented according to instructions in the supplemental material.
We discard Alignments and Overlapping whose value does not vary among our charts.
Besides, Emphasis and Flow are not considered since they are mainly concerned with key text or graphics, which are not well defined in charts.
We also combine those metrics (\textbf{All}).
Each metric consists of several features, which are fed into RankSVM to learn their weights.

\paragraph{Result}
\autoref{table:results} shows the results of the performances. 
In both experiments, 
our model outperforms the baseline RankSVM approaches.
In particular, RankSVM performs much poorer in Experiment 2,
showing that the relations between the impacts of each parameter on predicted scores tend to be non-linear.
All scoring baselines cannot achieve desired performances,
suggesting that those hand-crafted features for layout qualities in graphic design cannot be readily applicable to charts.


\subsection{Interpreting Models}
To understand the impact of layout parameters on the perceived layout quality,
we conduct the quantitative and qualitative analyses with the trained scoring model.
Those analyses help relate our work with prior knowledge about chart layout designs,
inform design guidelines, and provide qualitative support for our methods.
Specifically, we calculate the predicted layout quality score of different combinations of parameters. 
\begin{figure*}[!t]
    \setlength{\belowcaptionskip}{-5pt}
	\centering
	\includegraphics[width=1\linewidth]{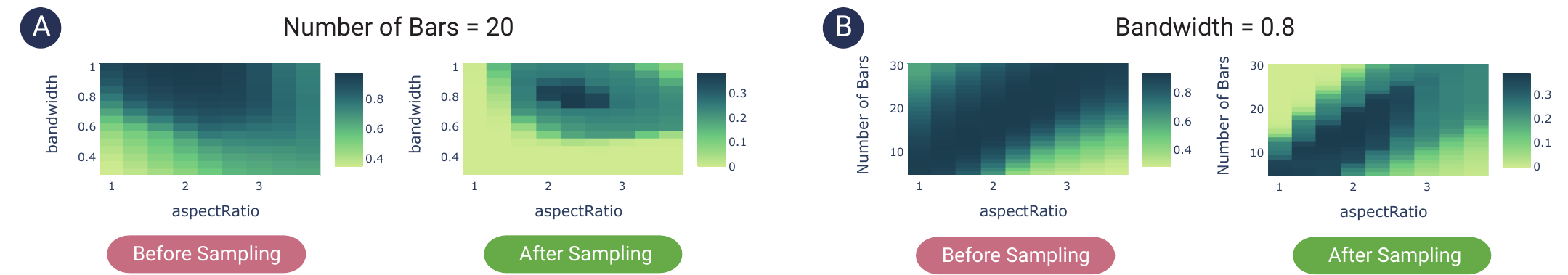}
	\caption{Heatmaps showing the predicted scores with different parameter combinations in Experiment 1. Our adaptive sampling strategies allow us to obtain fine-tuned results.}
	\label{fig:sampling}
    \Description[Visualizing the predicted score before and after sampling]{After sampling, the predicted score yields a small region of light colors (low scores) and a majority of dark colors (high scores).}
\end{figure*}

We study the relationships between the predicted score and each parameter by computing the correlations and visualizing distributions.
Our findings are summarized in the following text.
Those findings should be interpreted carefully since they are derived from the black-box ML models.
Thus, they should not be considered as guidelines, but instead speculative hypotheses that warrant future empirical research to confirm.

\subsubsection{Quantitative Parameter Analysis}
\autoref{table:corr} shows the Pearson correlations between the predicted scores and each parameter.
We first note the negative impacts of the number of bars in both experiments, 
showing that it is more challenging to find good layouts with more bars.
The aspect ratio contributes positively to the overall score, 
suggest that landscape layouts might be superior to portraits.

The impacts of bandwidths differ between our two experiments.
\autoref{fig:modelInterpret}{\protect\circled{A}} visualizes the predicted scores versus bandwidths using box-plots.
In Experiment 1, the average scores are relatively higher when the bandwidth is between 0.6 and 0.95, with a subtle peak at 0.8.
However, in Experiment 2 where horizontal bar charts are introduced,
the ``optimal'' interval become between 0.5 to 0.85, 
followed by a sharp drop after 0.9.
Based on those observations, we form hypotheses for future studies to confirm:
first, as a rule-of-thumb, the optimal bandwidth in vertical bar charts is 0.8;
second, the optimal bandwidth in horizontal bar charts is less than that in vertical bar charts.
Our second hypothesis conforms to existing rule-of-thumb that suggests a bandwidth between 0.57 to 0.67 in horizontal bar chart~\cite{Few2016}.

The label rotation has a moderate negative correlation (-0.43) with the score.
\autoref{fig:modelInterpret}{\protect\circled{B}} presents the combined effects of the label rotation with other parameters on the predicted score.
It is observed that non-rotation (zero-degree) is acceptable when the number of bars is small, when the aspect ratio is large, and when axis labels are short.
This is because axis labels are less likely to overlap with each other under those conditions.
Otherwise, the axis labels need to rotate to avoid overlapping.
In general, rotations by 45 degrees yield higher scores than rotations by 90 degrees.

We observe no correlation (0.04) between the orientation and the scores, showing that both horizontal and vertical bar charts have own advantages.
As shown in~\autoref{fig:modelInterpret}{\protect\circled{B}},
vertical bar charts achieve much fewer scores when the number of bars is large, when the aspect ratio is small than 1, and when the length of axis labels are large.
On the other hand, the scores of horizontal bar charts are less sensitive to the number of bars and the aspect ratio,
while horizontal charts seem strongly useful when axis labels are lengthy.

We also note that the score distributions are different between two experiments (\autoref{fig:modelInterpret}{\protect\circled{B}}).
The predicted scores in Experiment 1 vary between 0 to 0.39, while the range in Experiment 2 is 0.03 to 0.72.
This might be due to the existence of comparison ``deadlocks'' in Experiment 1, \eg{}, $\mathcal{I}_1 > \mathcal{I}_2, \mathcal{I}_2 > \mathcal{I}_3, \mathcal{I}_3 > \mathcal{I}_1$.
This left optimization constraints in~\autoref{equ:pairLoss}, \ie{},
$\mathcal{S}_1 - \mathcal{S}_2 > m, \mathcal{S}_2 - \mathcal{S}_3 > m, \mathcal{S}_3 - \mathcal{S}_1 > m$,
without feasible solutions.
In Experiment 1, we observe 21 three-node, 23 four-node, 10 five-node, and 32 six-node circles,
and the value of $m$ is 0.12.

\subsubsection{Qualitative Ablation Analysis}
We investigate the effectiveness of adaptive sampling strategies by conducting ablation analysis.
Specifically, we train two models on the dataset before and after the adaptive sampling process in Experiment 1, denoted \textbf{BS} and \textbf{AS}.
Two datasets are down-sampled to ensure the same data size.
\autoref{fig:sampling} presents two qualitative examples showing the learned relationships between the predicted scores and different parameters.
It is observed that the \textbf{BS} model yields a small region of light colors (low scores) and a majority of dark colors (high scores).
That said, it learns to reject bad conditions but could not further differentiate conditions scored ``borderline and above''.
On the contrary, the \textbf{AS} model is able to identify a small region of parameters that yield higher scores,
which accords with our goal to optimize layout parameters.
Besides, it identifies difficult conditions.
For instance,~\autoref{fig:sampling}{\protect\circled{B}} (right) 
suggests that the optimal aspect ratio increases with the number of bars before it reaches 3.
This implies the difficulties in finding good layout parameters for charts with an aspect ratio over 3,
which are uncommon and less favored.

\begin{figure*}[!t]
    \setlength{\belowcaptionskip}{-2pt}
	\centering
	\includegraphics[width=1\linewidth]{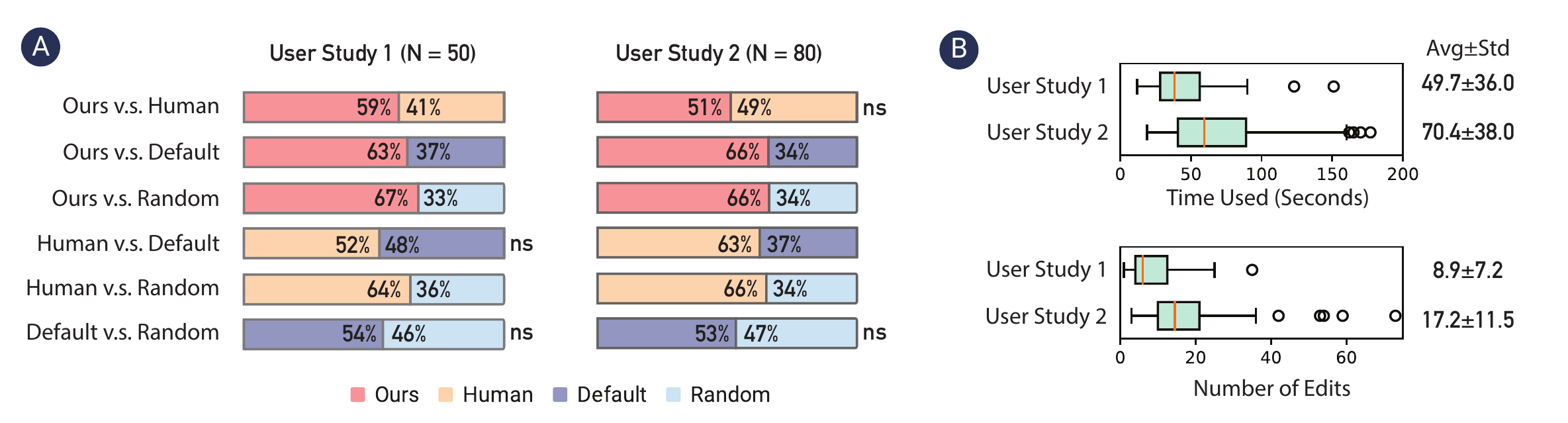}
	\caption{Results of the user study: {\protect\circled{A}} displays the results of group-wise comparison among four groups in terms of percentages of favored votes. An ``ns'' denotes no statistical significance via Wilcoxon signed-rank tests; {\protect\circled{B}} presents two box-plots visualizing the time used and the number of edits by laypeople in configuring the chart layout.}
	\label{fig:UserStudy}
    \Description[Results of the user study]{In User Study 1 (US1), our method outperforms Human, Default, and Random, and users on average spend 49.7 seconds on 8.9 times of edits. In User Study 2 (US2), both Ours and Human outperform Default and Random, while Default and Random are evenly matched. Users on average spend 70.4 seconds on 38 times of edits }
\end{figure*}
\section{Application}
\label{application}
To demonstrate the usefulness of \name{},
we present a novel application, \ie{}, automatic optimization of layouts.
Existing charting tools typically generate layouts by predefined heuristics,
which requires tedious manual adjustments. 
It would therefore be useful to automate this process by recommending layout parameters that improve the quality.
To that end, we propose an automatic optimization approach and conduct two user studies.

\subsection{Method}
We present two user studies in line with our experiments.
In User Study 1, 
we propose a common real-world scenario - presentations,
where individuals usually wish to create charts to convey data insights in an aesthetically pleasing manner.
The task is to adjust the aspect ratio and the bandwidth given the data.
User Study 2 concerns adaptive visualization designs, 
where a maximal width is posed as a hard constraint and the task is to adjust four parameters including the aspect ratio, the bandwidth, the orientation, and the label rotation.
We create 50 and 80 design cases for two studies, respectively, 
each case encoding randomly chosen data.
We compare our results (\textbf{Ours}) with those generated by laypeople (\textbf{Human}) and default parameters (\textbf{Default}), and random values (\textbf{Random}).

\paragraph{Our Approach}
Our optimization approach aims to find parameter values that maximize the layout quality score predicted by \name{}.
For that purpose, we adopt a brute-force method that enumerates combinations of values and selects the one with the highest predicted score.
\begin{rev}
We choose brute-force methods since the maximal enumeration size is 87,360 which computers could operate within seconds.  
Advanced optimization techniques are desired to cope with the expanding parameter space by avoiding enumeration~\cite{amaran2016simulation}.
\end{rev}

\paragraph{Human Baseline}
To obtain human baselines, we run an experiment on \MTurk{}. 
Participants are instructed to ``adjust the parameters until you are mostly satisfied with the layout''
with a What You See Is What You Get (WYSIWYG) editor implemented with Vega-Lite.
Akin to standard charting tools, participants are provided with a slider and an input box for adjusting continuous values, 
and a radio group for editing discrete parameters.
We record their editing history and the time used from the first editing to final submission.
Each participant is assigned to one and only one design task.

\paragraph{Default Baseline}
In User Study 1, we choose Microsoft Excel as the default baseline.
In order to keep the comparison fair,
we remove components that do not exist under other conditions, which include the chart title and y-axis gridlines.
Besides, the bars are filled with the default color of Vega-Lite.
In User Study 2, we compare our method against the Responsive Bar Chart feature in Vega-Lite\footnote{\url{https://vega.github.io/vega-lite/examples/bar_size_responsive.html}}.

\paragraph{Random Baseline}
The random baseline takes random parameters, which are sampled from values observed in the training data to make the comparison fair.

\subsection{Evaluation and Results}
We run another \MTurk{} experiment, asking participants to compare the results among the above four groups.
Similar to the labeling process,
we conduct pair-wise comparisons between every two groups.
Each between-group comparison includes 50 paired charts in User Study 1 and 80 in User Study 2.
Each paired chart is evaluated in a two-alternative force decision (2AFC) paradigm by 10 participants.
There is one duplicate pair out of per 10 for quality control.

\autoref{fig:UserStudy}{\protect\circled{A}} summarizes the results of group-wise comparisons in terms of the percentages of preferred votes in the 2AFC procedures.
In User Study 1 (US1), our method outperforms Human, Default, and Random ($p<0.05$, Wilcoxon signed-rank tests). 
It is also noted that while Human has a higher preference than Default,
the difference is not statistically significant.
Similarly, Default has a small yet not significant superiority over Random.
However, Human performs significantly better than Random.
This said, laypeople could only achieve a relatively small improvement in the layout quality, 
although they spent notable efforts, 
\ie{}, 49.7 seconds and 8.9 adjustments on average (\autoref{fig:UserStudy}{\protect\circled{B}}).

In User Study 2 (US2), both Ours and Human outperform Default and Random ($p<0.05$), while Default and Random are evenly matched.
This shows while Vega-Lite enables automatic adaptive visualization, the generated layouts are sub-optimal.
Our method only achieves compatible results with Human,
which might due to three reasons.
First, participants have spent more efforts,
\ie{}, 70.4 seconds and 17.2 adjustments on average (\autoref{fig:UserStudy}{\protect\circled{B}}).
It is therefore expected that participants could achieve better results.
Second, US2 presents a much more challenging task than US1,
since the total number of possible parameter combinations is 1,575 in US1 and 87,360 in US2 (\autoref{fig:parameter}).
However, the training data size is only 1,333 in US2,
which is far from fully representing the whole design space and therefore could not find the ``optimal'' solution all the time.
Still, our results are positive as our method has achieved human-level performances by leveraging a small training data,
showing the effectiveness of our sampling strategies.
Future work could further extend our work by augmenting the training data.
\begin{rev}
Finally, our sample size (80) is relatively small, considering the variety of the parameter space and the random nature in task generation.
In the future, we plan to conduct larger-scale user studies to better understand the scalability.
\end{rev}

In summary, our results show that the default heuristics for generating layouts in existing charting tools could result in sub-optimal results.
To improve the layout quality, laypeople need to engage in a time-consuming process to adjust the parameters over and again.
Our automatic approach could achieve at least human-level performance via small-sample learning, 
while removing the heavy costs of manual adjustments.

\section{Discussion and Conclusion}
We reflect on the implications and future work of our research.

\subsection{Implication}
\paragraph{Do not trust the defaults}
Charting tools and libraries provide default settings for user-configurable parameters.
Default settings are proven to introduce a default effect that people would blindly trust and stick with them~\cite{dinner2011partitioning}.
However, default settings are designed to be reasonable under most cases, \ie{}, to prevent stupid mistakes.
Thus, they are just acceptable but not good for all. 
We provide empirical evidence that the default layout parameters for bar charts in Excel and Vega-Lite are sub-optimal, 
which can be significantly improved by manual or automatic fine-tuning.
Those results support the needs of increasing recognition for utilizing default values prudently.

\paragraph{Augmenting empirical studies with a machine learning approach}
Our experiments can be considered as empirical studies aiming to identify the ``best'' combinations of variables.
This is challenging due to the vast design space, \ie{}, 87,360 possible combinations, making exhaustive enumeration and controlled studies infeasible.
In response, we propose a ML approach that learns to rank the combinations from small samples (1,333 pairs), yielding notable results.
More importantly, 
we formulate hypothesises of optimal variables via interpreting the ML model.
Future work could verify those hypothesises by conducting controlled experiments.

\paragraph{Quantifying visualizations with subjective metrics}
Recent years have witnessed a growing research interest in quantifying and benchmarking visualizations for machine learning (e.g.~\cite{hu2019viznet, saket2018task}).
Those work has predominately focused on objective metrics such as accuracy and effectiveness.
Subjective metrics, however, are relatively neglected, while they are considered more challenging to measure.
Our work extends this line of research by benchmarking charts with subjective metrics, \ie{}, human preference over layouts, through crowdsourcing experiments.
We describe our procedures and strategies for quantifying subjective metrics, hoping to inform future research to measure and improve visualizations from more diverse perspectives, \eg{}, understandability~\cite{shu2020makes}.

\paragraph{Improving aesthetic qualities of visualizations from a data-driven perspective}
A good visualization consists of four necessary elements: information, story, goal, and beauty~\cite{McCandless09}. 
In a broader sense, our work addresses the beauty, that is, the aesthetic quality.
We propose a data-driven method to learn human preference for layouts, 
which outperforms hand-crafted layout metrics.
Our results demonstrate the promising research possibilities of understanding and improving the aesthetic quality via data-driven machine learning approaches.
This research direction is supported by real-life practical needs,
\ie{}, 
existing charting tools generate sub-optimal results,
while laypeople tend to rely on default values~\cite{dinner2011partitioning} or need to engage in a time-consuming process to tune the results until they are satisfied with the result ~(\autoref{fig:UserStudy}{\protect\circled{B}}).
These needs call for an increasing recognition for understanding what makes a chart visually appealing and proposing more advanced automatic methods for improving the aesthetic quality.

\subsection{Critical Reflection}
We assess the quality of charts by asking participants ``which do you prefer?''.
Compared with scoring a single chart,
this paired comparison method is easier for participants and yields more precise and consistent results.
As such, we see potentials of adopting it for various purposes in visualization research.
To better inform future research, we discuss our critical reflections on this method.

\paragraph{Combating decision paralysis}
Decision-makings are not always easy, especially when the differences between two charts are small.
It could cause analysis paralysis where individuals overthink the situation that makes decision-making ``paralyzed''~\cite{langley1995between}.
Subsequently, individuals tend to choose an arbitrary decision hesitantly~\cite{tsukida2011analyze}.
As shown in~\autoref{fig:UserStudy}{\protect\circled{B}},
some participants spent much more efforts on editing the parameters than the average,
showing that they seemed subject to analysis paralysis.
To alleviate this problem,
future research should propose more effective sampling strategies that avoid over-subtle differences between paired charts.
Besides, we might borrow the idea of agile methodologies in software engineering to overcome the anti-pattern of decision paralysis~\cite{carkenord2009seven}.
One promising approach in the context of empirical research is to set time limits for viewing visualizations and making decisions (\eg{},~\cite{harrison2015infographic}).

\paragraph{``Evils'' can attract}
Psychological studies reveal the physical attractiveness stereotype that people tend to assume ``what is beautiful is good''~\cite{dion1972beautiful}.
In the context of data visualizations, this is exemplified by chartjunk~\cite{tufte2001visual,few2011chartjunk},
where laypeople are attracted by elements that are visually appealing but usually at the expense of effectiveness.
This contributes to the paradigm in charting tools that ``compromise'' with such human preference.
For example, Google Sheets supports 3D pie charts despite their criticism by the visualization research community.
Google Sheets also offers a Smooth Line Chart that improves the aesthetics but compromises the integrity of the underlying numbers.
Future research should be aware of this trade-off when designing the experiment settings.

\paragraph{Incorporating crowdsourced opinions with expert knowledge}
\begin{rev}
Visualization researchers have increasingly leveraged crowdsourcing experiments for the sake of scalability and diversity.
However, crowdsourcing experiments face challenges such as reduced control in the assessment of participants’ capability that might harm the validity~\cite{borgo2018information}.
Besides, 
we observe disputes in crowdsourced opinions.
To that end, we envision that expert knowledge could help increase validity, resolve disputes, and reduce costs.
For instance, one might select expert-generated charts as positive and randomly-generated charts as negative in pair~\cite{qian2020retrieve}.
However, it is worthy noting that expert judgement could clash with crowdsourced opinions that warrants deeper investigation~\cite{kutlu2018crowd}.
\end{rev}

\subsection{Limitation and Future Work}
\paragraph{Balancing human preference and perceptional effectiveness}
\begin{rev}
Our work takes only the first step in improving the visual quality of data visualizations via a data-driven approach that learns from human preference.
In particular, we study six layout parameters in bar charts.
We do not conduct comprehension experiments to evaluate their effects on perceptional effectiveness,
because the effect size of layouts on perceptions is typically small in standard bar charts~\cite{talbot2014four, zhao2019neighborhood}.
How to balance human preference and perceptional effectiveness is a clear next step for future work.
This is critical because layouts have proven to impose more influences in some other charts (\eg{}, ~\cite{heer2006multi, heer2009sizing}).
A key challenge here is that human preference and perception should be measured conjunctively in order to obtain the training data for machine learning approaches.
\end{rev}

 
\paragraph{Moving towards an more adaptive approach}
Although we propose two sampling strategies that enable learning from small data sets (Small Sample Learning),
our model in Experiment 2 only achieves human-level performance.
This presents a significant challenge as the size of the design space grows exponentially with the increasing number of parameters.
Future work should propose advanced sampling approaches to improve effectiveness.
Recent research in online adaptive sampling~\cite{katharopoulos2018not} that automatically updates the sampling strategy during training is a promising method to address this problem.
An interesting research problem would be how to dynamically adjust the sampling probabilities during crowdsourcing experiments. 
\textcolor{black}{Moreover, we see research opportunities in leveraging the authoring provenance (\eg{}, the editing histories) to augment the training data and develop a ML model that adaptive recommends design suggestions based on the current configuration.}

\paragraph{Understanding the representations and models for visualization research}
\begin{rev}
In a broader sense, it remains an open challenge to choose the feature representations and machine learning models for visualizations.
Similar with Draco~\cite{moritz2018formalizing} and VizML~\cite{hu2019vizml},
\name{} is trained on the parameter features that are compact and computationally inexpensive,
which, however, might not generalize to unobserved parameter values (\eg{}, more than 30 bars) and different chart types or require labour-intensive feature engineering.
Although graphical features (\ie{}, bitmaps) might embrace generalisability,
recent studies~\cite{fu2019visualization,haehn2018evaluating} suggest that CNNs, the most common model for analyzing visual imagery~\cite{valueva2020application}, seem not currently capable of processing visualization images.
This underscores the research needs to explore advanced ML models, \eg{}, VAE~\cite{fu2019visualization}.
Furthermore,
\name{} does not include the underlying data distributions and non-layout parameters (\eg{}, colors) in the training representations, 
which could influence the perceived aesthetic qualities.
To that end, future research should study how to choose and fuse multiple representations including the underlying data, parameters, and graphics.
\end{rev} 

\paragraph{Debating ``what is beautiful is good''}
Finally, we propose a research agenda towards more understanding of the roles of aesthetic qualities in data visualizations.
This is critical since nowadays more and more people are able to create visualizations,
so does their exposure to the greater masses.
This phenomenon contributes to the increasingly popular pursuits of aesthetic qualities.
We even see extreme cases where aesthetic concerns play a more crucial role than usability and even usefulness, \eg{}, the Smooth Line Chart.
How should the research community respond to this shifting boundary?

\begin{acks}
The research is partially supported by Hong Kong RGC GRF grant 16213317. 
\end{acks}

\bibliographystyle{ACM-Reference-Format}
\bibliography{main}

\appendix

\end{document}